\documentclass[aip,apl,preprint,amsmath,floatfix,noeprint]{revtex4-2}

\usepackage{graphics}
\usepackage{natbib}
\usepackage{graphicx}
\usepackage{amsmath}
\usepackage{bm}
\usepackage{lineno}
\usepackage{color}
\usepackage{dcolumn}
\usepackage[utf8]{inputenc}
\usepackage[T1]{fontenc}
\usepackage{mathptmx}
\DeclareUnicodeCharacter{2215}{ }
\DeclareUnicodeCharacter{2212}{ }
\begin{document}
\title{Competing Uniaxial Anisotropies in  Epitaxial Fe Thin Films Grown on InAs(001)}
\author{James M. Etheridge}
\author{Joseph Dill}
\affiliation{School of Physics and Astronomy, University of Minnesota, Minneapolis, Minnesota 55455}
\author{Connor P. Dempsey}
\author{Mihir     Pendharkar}
\affiliation{Electrical and Computer Engineering, University of California Santa Barbara, Santa Barbara, California 93106, USA}
\author{Javier Garcia-Barriocanal}
\author{Guichuan Yu}

\affiliation{School of Physics and Astronomy, University of Minnesota, Minneapolis, Minnesota 55455}

\author{Vlad S. Pribiag}
\author{Paul A. Crowell}
\email{crowell@umn.edu}
\affiliation{School of Physics and Astronomy, University of Minnesota, Minneapolis, Minnesota 55455}

\author{Chris J. Palmstr{\o}m}

\affiliation{Electrical and Computer Engineering, University of California Santa Barbara, Santa Barbara, California 93106, USA}
\affiliation{Materials Department, University of California Santa Barbara, Santa Barbara, California 93106, USA}

\begin{abstract}
We report on the interplay of two uniaxial magnetic anisotropies in epitaxial Fe thin films of varying thickness grown on InAs(001) as observed in ferromagnetic resonance experiments. One anisotropy originates from the Fe/InAs interface while the other originates from in-plane shear strain resulting from the anisotropic relaxation of the Fe film. X-ray diffraction was used to measure the in-plane lattice constants of the Fe films, confirming the correlation between the onset of film relaxation and the corresponding shear strain inferred from ferromagnetic resonance data. These results are relevant for ongoing efforts to develop spintronic and quantum devices utilizing large spin-orbit coupling in III-V semiconductors.
\end{abstract}
\date{\today}
\maketitle

Ferromagnetic thin films grown epitaxially on semiconductor substrates have been of interest in the field of spintronics for several decades. They serve as model systems for spin injection and spin detection, which would be critical components of a spin transistor.\cite{Datta1990, Sato2001, Hanbicki2002}  For example, the system of Fe on GaAs(001) has been extensively studied due to the near lattice match of GaAs and Fe as well as the ability to form Schottky tunnel barriers.\cite{Hanbicki2002, Thomas2003, Lu2009, Lou2007, Christie2015, LeBeau2008} However, the introduction of semiconductors with larger spin orbit coupling (SOC) is necessary for the realization of a fast and efficient spin field-effect transistor. \cite{Manchon2015, Koo2009, Chuang2015} Recently, there has also been growing interest in developing quantum devices that integrate magnetic elements with large-SOC semiconductors. \cite{liu2019coherent, yang2020_spinTransport, sun2020_spinFiltering, vaitiekenas2021zero, yuDFT2021, Yang2021} The two most commonly used  III-V semiconductors with large SOC are InAs and InSb, whose lattice mismatch with Fe are 5.4\% and 11.5\%, respectively. This is quite large compared to the 1.4\% mismatch between Fe and GaAs. Despite the possible challenges that a large lattice mismatch may present, ferromagnets (FM) on III-V semiconductors with strong SOC must be investigated further given their potential for applications.

When epitaxial ferromagnets with a bulk cubic symmetry are grown on III-V surfaces, there are two primary sources of in-plane uniaxial magnetic anisotropy. The first arises from the anisotropic interfacial bonding between the substrate and FM film due to the breaking of the fourfold symmetry at the III-V surface. Independent of the III-V surface reconstruction, the easy axis of this anisotropy is along the [110] direction.\cite{Kneedler1997, Herfort2004, Moosbuhler2002, Brockmann1999, Freeland2001} The second uniaxial anisotropy is due to a magnetoelastic energy originating from anisotropic in-plane strain.\cite{Thomas2003, Lu2009a} This results in a magnetoelastic energy with a minimum along the [1$\bar{1}$0] direction. The end result is a thickness-dependent competition between these two uniaxial terms as well as the FM bulk anisotropy. In this Letter we demonstrate the FM thickness dependence (1.4 to 39.0 nm) of the competition between the two uniaxial anisotropies as well as the cubic anisotropy of bulk Fe, whose easy axes are along the <100> direction of Fe/InAs(001) heterostructures. To probe these anisotropies room temperature broadband ferromagnetic resonance (FMR) measurements were performed. In addition to the FMR experiments, reciprocal space mapping (RSM) was performed using x-ray diffraction (XRD) to determine lattice constants and subsequent lattice strain\cite{Yu2021}. %(add Guichan citation for RSM)

The Fe films  were grown by molecular beam epitaxy (MBE). The InAs(001) substrates were cleaned for one hour in an ultrahigh vacuum (UHV) chamber with a base pressure $\sim$ $1\times10^{-10}$ mbar  using atomic hydrogen. The temperature of the substrate was 475 $^{\circ}$C, as read by a thermocouple close to the sample, and the hydrogen cracker was  1700 $^{\circ}$C. After cleaning, the substrates were loaded into a VG V80 UHV growth chamber with a base pressure $\sim$ $5\times10^{-11}$ mbar, where elemental Fe was deposited on the substrate from an effusion cell. During deposition the substrate was maintained at room temperature except for the 4.6 nm film, where the sample was heated to 75$^{\circ}$C as measured by a thermocouple. After the deposition of Fe was completed, the samples were either capped with approximately 6 nm of Ti in the same VG V80 chamber as the Fe, or \textit{in vacuo} transferred to an inter-connected UHV chamber where a 10 nm Au cap was grown at room temperature. The lower growth temperatures used in this work were chosen to avoid diffusion of Fe into InAs and to minimize any interfacial reactions.

Reflection high energy electron diffraction (RHEED) performed during growth confirmed the film was oriented as expected as well as suggesting a single crystalline phase. Post-growth, grazing incidence x-ray diffraction (GIXRD) was used to determine the precise Fe thickness for each sample. Vibrating sample magnetometry yielded hysteresis loops for each sample at multiple orientations, enabling the determination of the saturation magnetization and coercivity for each sample. Using the Stoner-Wolfarth model, theoretical hysteresis loops were generated with results qualitatively agreeing with the experimental results. Wide-angle XRD was performed in concert with the RSM, with both pointing to the same results, which are discussed in detail below.
\begin{figure}
\centering
\includegraphics{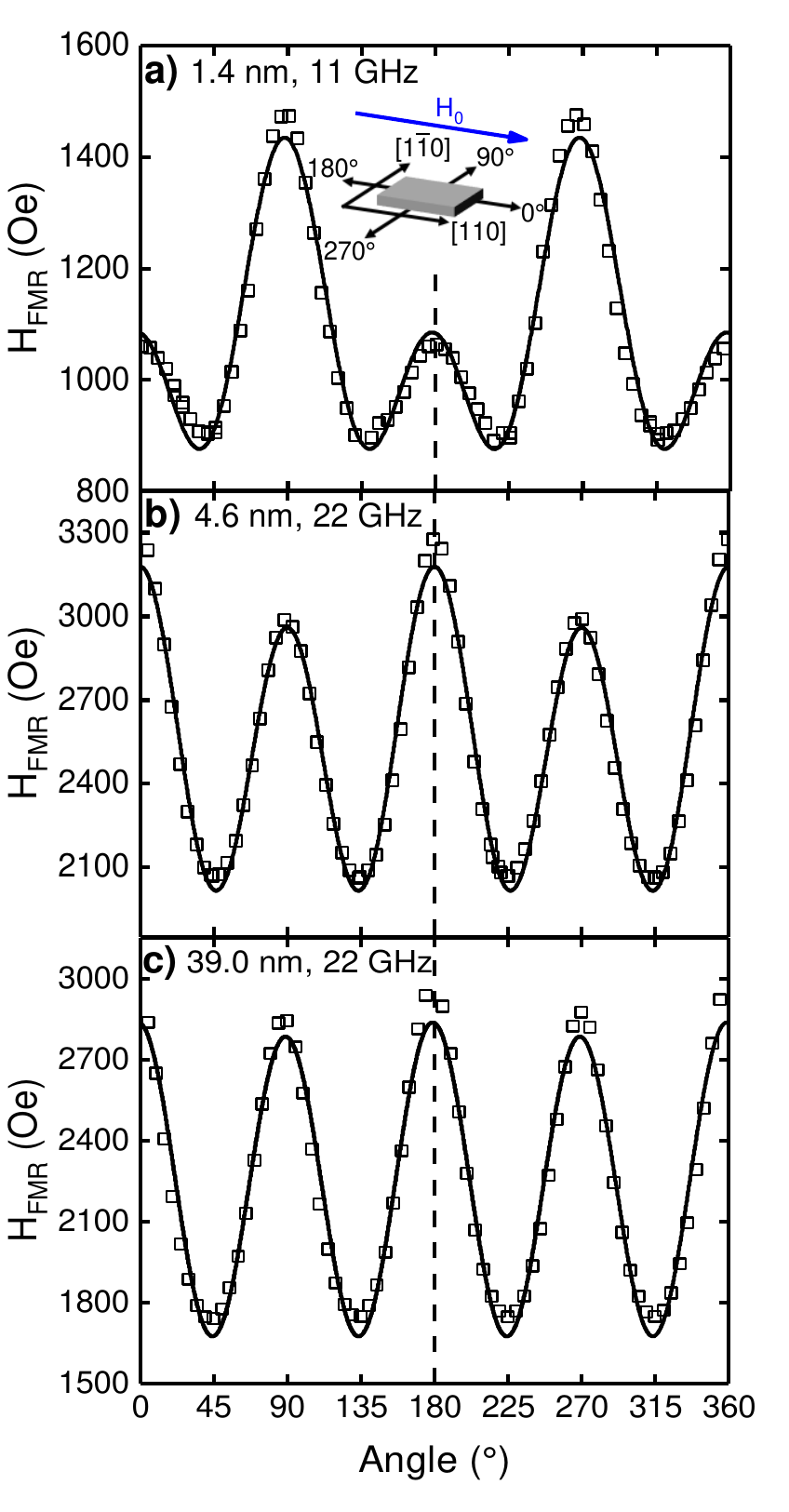}
\caption{a) The inset shows the geometry of the sample with respect to the applied magnetic field. a)-c) Azimuthal angular dependence of resonance fields with corresponding fits to Eq. \ref{suhl} shown in the solid curves.}
\label{FMR 195}
\end{figure}

Broadband FMR measurements were performed at room temperature using a coplanar waveguide (CPW) in a simple transmission geometry with modulation of the applied magnetic field.\cite {Peria2020} The applied field was swept through resonance, and the signal was fit to a Lorentzian derivative function. From these fits $H_{FMR}$, the strength of the applied field at resonance, was extracted. This procedure was repeated as $\phi_H$, the angle of the applied field in the plane of the film, was varied from 0$^{\circ}$ to 360$^{\circ}$ in increments of 5$^{\circ}$. The data for several samples are shown in Fig.  \ref{FMR 195}. They show a combination of a two-fold and four-fold symmetry of varying strengths dependent on the sample thickness. For the thinnest sample, the two-fold component is strongest with respect to the four-fold component. The magnetic easy axis of this sample is along the [110] direction of the crystal. For the intermediate thickness of 4.6 nm, the easy axis switches to the [1$\bar{1}$0] direction, and there is an increase in the relative strength of the four-fold component. The 90$^{\circ}$ rotation of the easy axis signals the relaxation of the Fe film, and this will be discussed in depth later. In the thicker samples, the magnetic easy axis is along the [1$\bar{1}$0] direction, indicating that they have also relaxed. When the film thickness reaches 39.0 nm, the data show a dominant four-fold symmetry that would be expected from a bulk Fe sample. However, a small two-fold component, whose easy axis is along the [1$\bar{1}$0] direction,  still remains, suggesting that complete relaxation has not yet occurred.

To quantitatively describe the magnetic anisotropy of this series of heterostructures, a model of the free energy density was introduced with Zeeman, two uniaxial, cubic, and shape anisotropy contributions,
\begin{equation}\label{Free Energy Density}
\begin{split}
F = &-\mathbf{M}\cdot\mathbf{H}+K_u\, \sin^{2}\phi+K_{u2}\,\sin^{2}(\phi+\pi/2) \\
& +K_1(\alpha_1^{2}\alpha_2^{2}+\alpha_2^{2}\alpha_3^{2}+\alpha_3^{2}\alpha_1^{2})+2\pi M_{eff}^{2}\,\cos^{2}\theta,
\end{split}
\end{equation}
where $\mathbf{M}$ is the magnetization vector, $\mathbf{H}$ is the applied magnetic field vector, $K_u$ is the uniaxial anisotropy constant originating from the Fe/InAs interface, $K_{u2}$ is the uniaxial anisotropy constant originating from anisotropic strain, $K_1$ is the cubic anisotropy constant, $\alpha$'s are the directional cosines of the magnetization, and $M_{eff}$ is the effective magnetization. The concept of effective magnetization comes from the combination of demagnetizing energy and perpendicular magnetic anisotropy. This free energy density can be used to determine the condition for FMR, first shown by Suhl,\cite{Suhl1955} with the result being,
\begin{equation}\label{suhl}
\omega = \frac{\gamma}{M_s \sin{\theta}} \sqrt{F_{\theta\theta} F_{\phi\phi} - \left(F_{\theta\phi}\right)^2,}
\end{equation}
where $F$ is the free energy density, $\omega$ is the frequency, $\gamma$ is the gyromagnetic ratio, $M_s$ is the saturation magnetization, $\phi$ is the azimuthal angle of the magnetization with respect to the [110] direction, and $\theta$ is the polar angle of the magnetization with respect to the [001] direction. The subscripts denote derivatives with respect to $\theta$ and $\phi$.

We made the ansatz that $K_u$ is a 3D effective surface energy density, so it must be inversely proportional to film thickness.\cite{Lu2009a,Moosbuhler2002,Brockmann1999} Thus,
\begin{equation}\label{Ku}
K_u = \frac{K_u^{int}}{t},
\end{equation}
where $t$ is the thickness of the film and $K_u^{int}$ is the purely surface portion of the uniaxial anisotropy. It is assumed that $\mathbf{M}$ and $\mathbf{H}$ are in the plane of the film. A previously reported value of 1714  $\textrm{emu}/\textrm{cm}^{3}$ was used for the magnetization.\cite{Cullity2008} At resonance, the applied magnetic field is large enough, greater than 800 Oe, to assume that the magnetization is in the same direction as the applied field. This assumption was verified for all films through hysteresis loops, including measurements with the applied field along the magnetic hard axis. Eq. \ref{suhl} was manipulated to express $H_{FMR}$ as a function of $\phi_H$ in order to fit the experimental data of $H_{FMR}$ versus $\phi_H$. From these fits, $K_u$, $K_{u2}$, $K_1$, and $M_{eff}$ were extracted.

\begin{figure}
\centering
\includegraphics{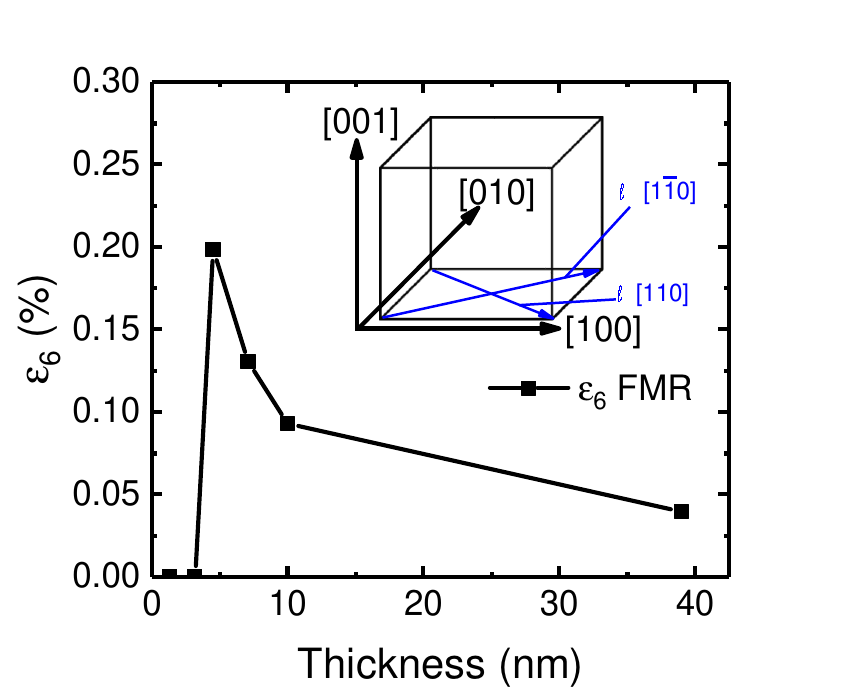}
\caption{The inset shows a parallelepiped and its in-plane diagonals. Shear strain is plotted as a function of Fe thickness with values determined from fitting FMR data to Eq. \ref{suhl}.}
\label{Shear Strain}
\end{figure}

\begin{figure}[h!]
\centering
\includegraphics{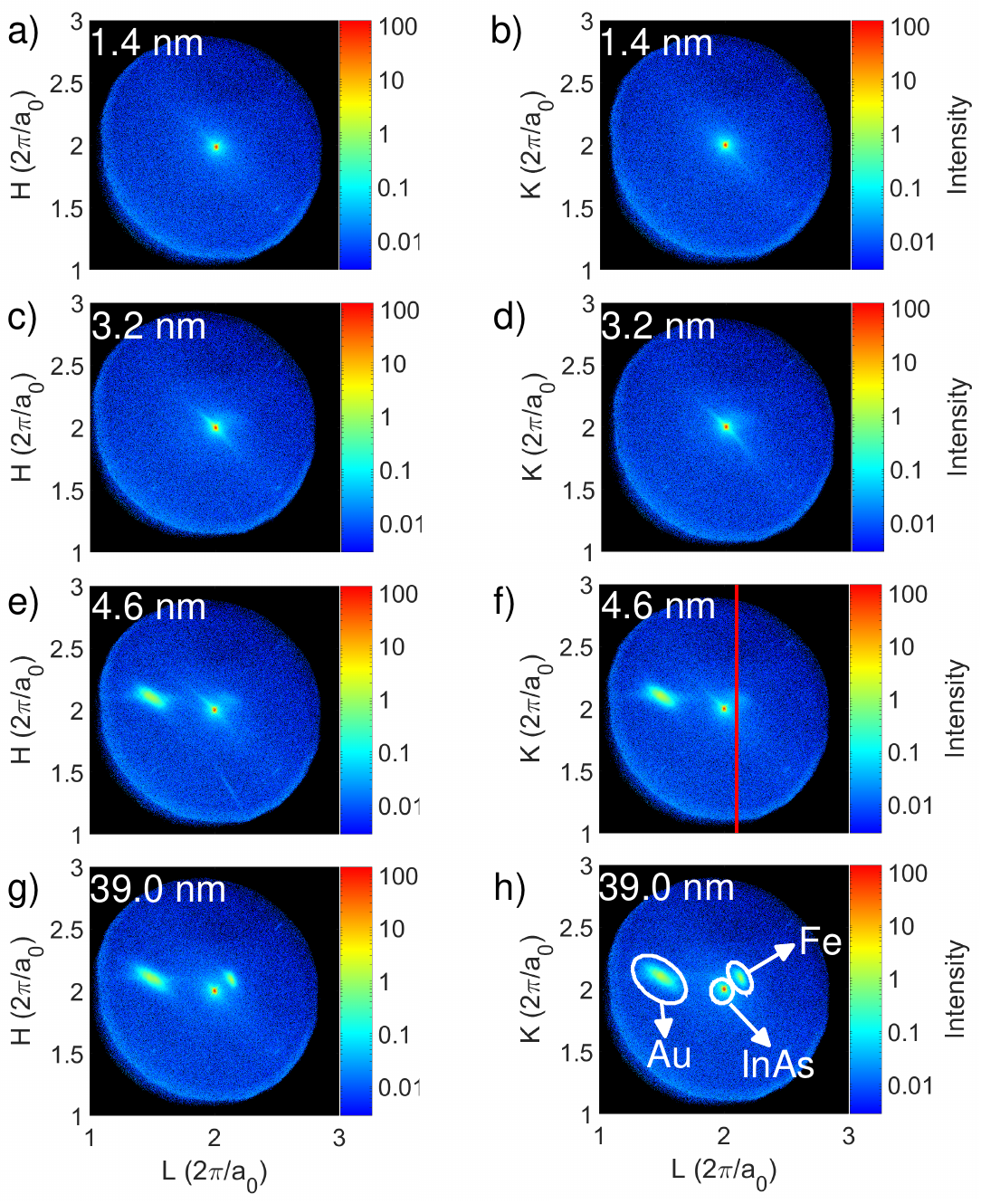}
\caption{a) c) e) g) Reciprocal space maps around the (202) peak of InAs(001). b) d) f) h) Reciprocal space maps around the (022) peak of InAs(001). The Au peak seen in panels e)-h) comes from the Au capping layer. This peak is not seen in a)-d) as the 1.4 and 3.2 nm samples were capped with Ti. The units are Miller indices of InAs, with a\textsubscript{0} = 6.06 \AA. The red line in f) corresponds to the projection of the RSM in Fig. \ref{Gaussian Fit}.}
\label{XRay 195}
\end{figure}

For the thinnest sample, 1.4 nm, the film is pseudomorphic, so $K_{u2}$ can be set to zero. This determination was made in light of the following results: for thicker samples, greater than 3.2 nm, the easy axis is rotated 90$^{\circ}$, and an Fe peak appears in the x-ray diffraction data. A pseudomorphic sample has an Fe peak on top of the InAs peak making it very difficult to separate the two. The remaining free parameters were determined through fitting. With $K_u$ extracted and the thicknesses determined through x-ray reflectivity (XRR), $K_u^{int}$ was determined to be $2\times10^{-2} \textrm{erg}/\textrm{cm}^{2}$. Previous works on Fe/GaAs heterostructures have seen this surface energy term $\sim 10^{-1} \textrm{erg}/\textrm{cm}^{2}$.\cite{Moosbuhler2002,Brockmann1999} With all thicknesses known and $K_u^{int}$ extracted, $K_u$ was constrained for subsequent fits. The data from the 3.2 nm film are more ambiguous than that of the 1.4 nm film. The magnetic easy axis of the 3.2 nm sample is along the [110] direction of the crystal, just as it is in the 1.4 nm sample. In the x-ray diffraction data, seen in panel c) and d) of Fig. \ref{XRay 195}, there is a shoulder of the InAs peak which is most likely the emergence of an Fe peak. However, $K_{u2}$ must still be small compared to $K_{u1}$ as there is not a 90$^{\circ}$ rotation of the easy axis. Thus, $K_{u2}$ again was set to zero for its corresponding fit. None of the other samples were pseudomorphic and the x-ray data were readily analysable, so $K_{u2}$ was not constrained for their fits.
\begin{figure}
\centering
\includegraphics{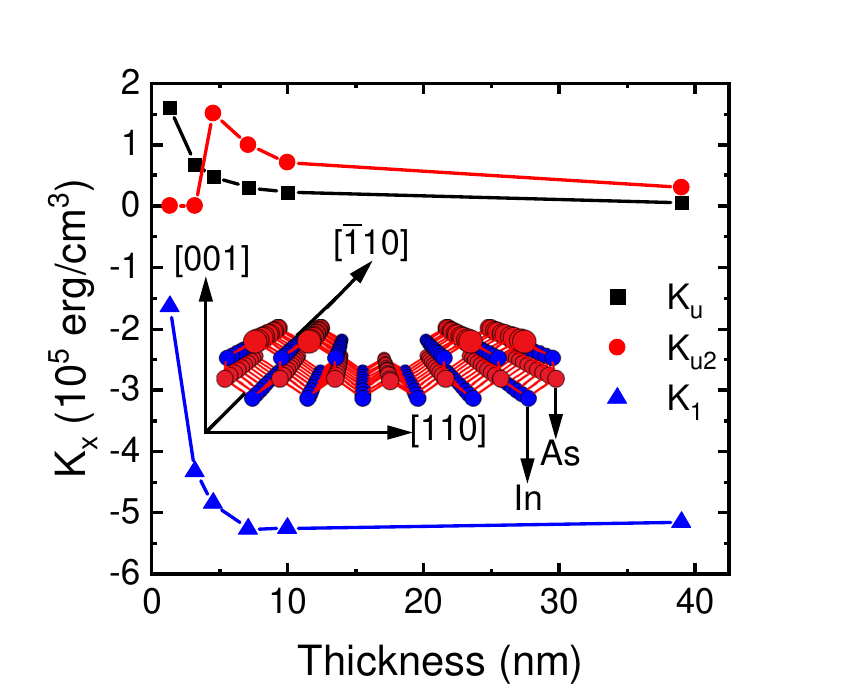}
\caption{The inset shows the structure of the InAs(001) including the direction of the As dimer rows. Anisotropy constants are plotted as a function of Fe thickness.}
\label{Anisotropy Constants}
\end{figure}
The values of $K_u$, $K_1$, and $K_{u2}$ resulting from this procedure are shown in Fig. \ref{Anisotropy Constants}. By the above ansatz, the magnitude of $K_u$ is inversely proportional to thickness as defined in Eq. \ref{Ku}. The magnitude of $K_1$ increases significantly with thickness, saturating for thicknesses above 7.1 nm to a value of $5.2\times10^{5} \textrm{erg}/\textrm{cm}^{3}$. This is slightly larger than the literature value for bulk Fe at room temperature, $4.8\times10^{5} \textrm{erg}/\textrm{cm}^{3}$.\cite{Cullity2008} $K_{u2}$ is zero for the two thinnest samples and then also decreases with thickness. The appearance of a non-zero $K_{u2}$ signals that the Fe has relaxed, albeit not completely, resulting in an in-plane shear strain of the Fe lattice. From the definition of magnetoelastic energy,\cite{Paes2013, Hoselitz1962} $K_{u2}$ must take the following form,
\begin{equation}\label{Ku2}
K_{u2} = B_2 \epsilon_6,
\end{equation}
where $B_2$ is the magnetoelastic coefficient of Fe associated with shear strain, and $\epsilon_6$ is the in-plane shear strain defined as $\epsilon_{[110]}-\epsilon_{[1\bar{1}0]}$. The extracted values of $\epsilon_6$, shown in Fig. \ref{Shear Strain}, are consistent with values from previous Fe/GaAs heterostructures, which are on the order of 0.1\%.\cite{Thomas2003,Lu2009} For the set of samples studied, the critical thickness at which shear strain is introduced to the system occurs between 3.2 and 4.6 nm. After the relaxation of the film and subsequent onset of shear strain, the shear strain decreases with thickness, but does not reach zero. This indicates that the film is not completely relaxed even at a thickness of 39.0 nm.

 We attempted to determine $\epsilon_6$ directly through XRD and compare to the value extracted from the FMR data. To do this the lattice constants of each sample where determined through reciprocal space mapping with several results shown in Fig. \ref{XRay 195}. Data taken for two different geometrical configurations offset by a 90$^{\circ}$ rotation of the sample about the [001] direction. This was done so that maps around both the (202) and (022) substrate peaks could be produced. From the RSM's along with the FMR results, it is determined that the thinnest sample is pseduomorphic to the substrate. The data of the second thinnest sample has a shoulder on the InAs peak, most likely corresponding to an Fe peak. Due to low signal to noise lattice constants could not be extracted. Despite this, the FMR data for this sample points to a small and possibly negligible shear strain. The appearance of a definite offset Fe peak in the thicker samples shows that the Fe layer has relaxed. To determine the amount of relaxation, the RSM's, oriented by the substrate peak, were fit to Gaussian curves and the effective Miller indices were extracted. With the effective Miller indices of the (101) and (011) Fe peaks known, the lattice constants of each orientation were calculated. The map centered on the (202) peak gives the \textit{a} and \textit{c} lattice constants while the one centered on the (022) peak gives \textit{b} and \textit{c}. Lattice parameter values are shown in Table \ref{Lattice Paramaters}. The \textit{c} parameter determined by wide angle XRD in a coupled geometry was consistent with the results of the RSM's, confirming the validity of the procedure.

\begin{table}
    \caption{Summary of the lattice parameters determined from RSM's shown in Fig. \ref{XRay 195}.}
    \label{Lattice Paramaters}
    \begin{ruledtabular}
    \begin{tabular}{l l l l l}
       Fe Thickness (nm) & a (\AA) & b (\AA) & c (\AA) & a\textsubscript{InAs}/2 (\AA) \\
        \hline
        4.6 & 2.908 & 2.908 & 2.868 & 3.03\\
        7.1 & 2.906 & 2.907 & 2.840 & 3.03\\
        10.0 & 2.902 & 2.901 & 2.836 & 3.03\\
        39.0 & 2.892 & 2.892 & 2.839 & 3.03\\
    \end{tabular}
    \end{ruledtabular}
\end{table}

\begin{figure}[h!]
\centering
\includegraphics{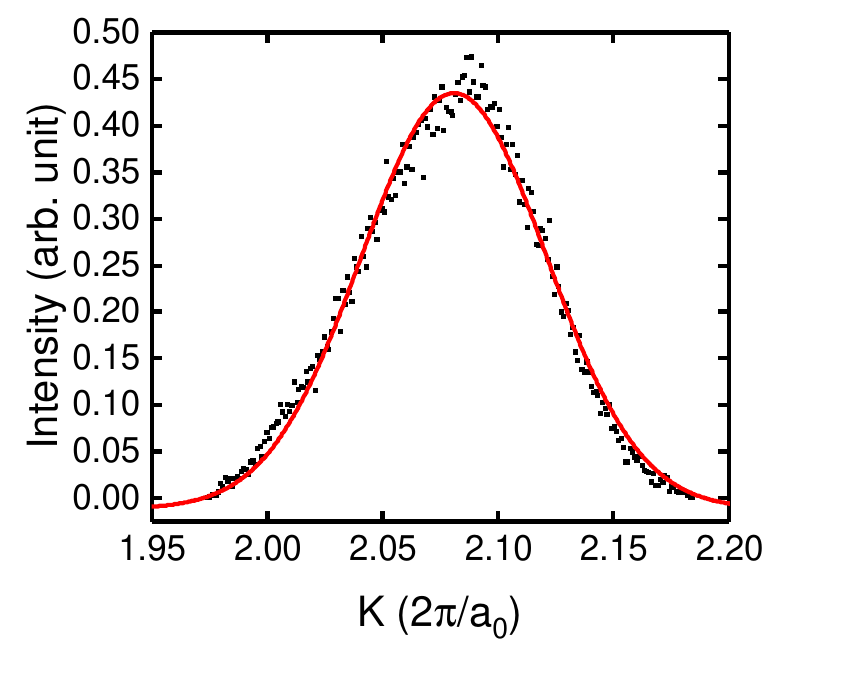}
\caption{An example of Gaussian fitting, used to extract the effective Miller indices, of one of the Fe peak projections of the 4.6 nm sample shown as the red line in panel f).}
\label{Gaussian Fit}
\end{figure}

Like the FMR data, the RSM's show that films thicker than 3.2 nm are relaxed. The data from the 3.2 nm sample is ambiguous in regards to if it pseudomorphic or not. However, even if the sample has relaxed, little to no shear strain is introduced to the lattice as inferred from the FMR data. For films 4.6 nm thick and greater, the lattice develops a tetragonal distortion, with equal \textit{a} and \textit{b} lattice parameters larger than the out-of-plane \textit{c} parameter. As the thickness of the film increases the in-plane lattice constants relax towards the bulk value of 2.866 \AA,  while maintaining the tetragonal distortion. It is inferred from the FMR data that the lattice develops decreasing amounts of in-plane shear strain with increasing thickness as expected. This strain distorts the cubic lattice into a parallelepiped. All angles defining the parallelepiped should be extremely close to but not exactly 90$^{\circ}$, resulting in a base that is nearly a square, but is in actuality a parallelogram whose diagonals differ in length. The shear strain, $\epsilon_6$, which is defined as the difference of the two diagonal strains, will be nonzero.  To determine the in-plane shear strain, both in-plane lattice parameters and the resulting diagonals must be determined from a single XRD scan, before rotating the sample 90$^{\circ}$. Unfortunately, from the XRD geometry used to acquire the data, this was not possible. A determination of the shear strain using a constant volume assumption along with the lattice parameters shown in Table \ref{Lattice Paramaters} was attempted. However, the resulting in-plane shear strains are orders of magnitude larger than those expected from the FMR data and thus non-physical. It is believed that the constant volume assumption is not valid in our samples. This is not surprising as Poisson's ratio for Fe is 0.29, and a material that perfectly adheres to the the constant volume assumption would have a Poisson's ratio of exactly 0.5.\cite{Ledbetter1973}

It is interesting to note that the 39.0 nm sample is not completely relaxed with \textit{a} and \textit{b} parameters larger than bulk and the \textit{c} parameter less than bulk. Although this is surprising, the FMR data also supports the conclusion that the 39.0 nm sample is not completely relaxed. Fig. \ref{Shear Strain} clearly shows a small nonzero shear strain. Although direct evidence of shear strain through in-plane x-ray diffraction was not investigated, the agreement between current FMR and x-ray data is strong evidence of its existence.

We have shown through FMR and XRD that the magnetic anisotropy of Fe/InAs(001) heterostructures is heavily dependent on sample thickness. Our investigations show that anisotropic relaxation of the Fe thin films results in shear strain, producing an additional term of magnetoelastic origin in the free energy density. This magnetoelastic term was confirmed through a rotation of the uniaxial easy axis as observed through FMR measurements. Similar phenomena have been reported in Fe/GaAs(001) structures. This suggests that, despite Fe/InAs having a significantly larger lattice mismatch than Fe/GaAs, this does not appear to have a detectable detrimental effect on interfacial magnetic properties. We would expect to see similar results for other Fe/III-V semiconductor heterostructures as well, which is encouraging for the developement of applications relying on the interplay between ferromagnetism and high-SOC semiconductors.

\begin{acknowledgments}
We acknowledge Bill Peria and Brett Heischmidt for their helpful input and discussions throughout this research. This work was supported by the Department of Energy under award no. DE-SC0019274. Parts of this work were carried out in the Characterization Facility, University of Minnesota, which receives partial support from NSF through the MRSEC program.
\end{acknowledgments}

%

%\bibliography{Library_Etheridge}

\end{document}